\documentclass[12pt,notitlepage]{article}

\usepackage{amsfonts}
\usepackage[nosumlimits]{amsmath}
\usepackage{amssymb}
\usepackage{amsthm}

\usepackage{feynmf}
\usepackage{latexsym}
\usepackage{mathrsfs} 

\allowdisplaybreaks

\newcommand{\dd}[2]
{\frac{\delta {#1}}{\delta {#2}}}

\newcommand{\pp}[2]
{\frac{\partial{#1}}{\partial{#2}}}

\newcommand{\boxx}[1]         
{\boxed{\;{#1}\;}}

\newcommand{\boxml}[1]        
{\fbox{\addtolength{\linewidth}{-11\fboxsep}
\begin{minipage}[t]{\linewidth}\vspace{-18pt}
\bml #1 \nn\end{multline}
\end{minipage}}\\[9pt]}

\newcommand{\boxea}[1]        
{\bce\fbox{
    \begin{Beqnarray}
    \;#1\;
    \end{Beqnarray}}\ece}

\newcommand{\LL}{[\![} \newcommand{\RR}{]\!]}

\def\ts{\mspace{1.5mu}}

\newcommand{\nc}\newcommand
\nc{\rnc}\renewcommand

\nc{\ph}\indent                \nc{\flush}\noindent
\nc{\bfr}{\begin{flushright}}    \nc{\efr}{\end{flushright}}
\nc{\bce}{\begin{center}}        \nc{\ece}{\end{center}}
\nc{\bq}{\begin{quote}}        \nc{\eq}{\end{quote}}
\nc{\vs}{\vspace}              \nc{\ms}{\mspace}

\nc{\nn}\nonumber
\nc{\bay}{\begin{array}}     \nc{\eay}{\end{array}}
\nc{\bal}{\begin{align}}    
\nc{\bcs}{\begin{cases}}     \nc{\ecs}{\end{cases}}      
\nc{\beq}{\begin{equation}}  \nc{\eeq}{\end{equation}}
\nc{\bea}{\begin{eqnarray}}  \nc{\eea}{\end{eqnarray}}
\nc{\bmx}{\begin{matrix}}   
\nc{\bml}{\begin{multline}} 
\nc{\bp}{\begin{proof}}      \nc{\ep}{\end{proof}}       
\nc{\bpmx}{\begin{pmatrix}}  \nc{\epmx}{\end{pmatrix}}
\nc{\bsp}{\begin{split}}    
\nc{\beqs}{\beq\bsp}        

                       \nc{\bs}\boldsymbol
\nc{\tbf}\textbf       \nc{\mbf}\mathbf   
                       \nc{\mfk}\mathfrak 
\nc{\trm}\textrm       \nc{\mrm}\mathrm   
                       \nc{\mcl}\mathcal  
                       \nc{\mrs}\mathscr  
\nc{\tsl}\textsl
                       \nc{\mbb}\mathbb   
\nc{\tit}\textit       \rnc{\mit}\mathit  
\nc{\ttt}\texttt       \nc{\mtt}\mathtt   
                       \nc{\mnl}\mathnormal
\nc{\tx}\text
\nc{\disp}\displaystyle

\rnc{\(}{\left(}         \rnc{\[}{\left[}
\rnc{\)}{\right)}        \rnc{\]}{\right]}
\rnc{\>}{\;\,}           \nc{\spc}{\qquad\qquad}
\nc{\2}{\;\,\;\,}        \nc{\bspc}{\qquad\qquad\qquad\qquad}
\nc{\3}{\;\,\;\,\;\,}    \nc{\<}{\!\!\!\!\!}
\nc{\ee}{\!\!\!=\!\!\!}
\def\bra{\langle}        \def\ket{\rangle}

\nc{\lh}{\lrcorner\,}

     \def\oset{\overset}

\def\al{\alpha}
\def\be{\beta}       \def\de{\delta}
\def\gam{\gamma}     \def\Gam{\Gamma}    
\def\eps{\epsilon}   \def\veps{\varepsilon}
\def\lam{\lambda }      
                     \def\vph{\varphi}   
                                         \def\vPsi{\varPsi}

                     \def\Sig{\Sigma}    
\def\om{\omega}      \def\Om{\Omega}     
\def\tha{\theta}

\def\ra{\rightarrow}        
\def\lgra{\longrightarrow}

\def\Ra{\Rightarrow}        

\def\sea{\searrow}          
\def\vgeq{\geqslant}        \def\vleq{\leqslant}
\def\p{\partial}            

                \nc{\bx}{\square}
                \nc{\bt}{\bullet}
\def\bcdot{\boldsymbol\cdot}
\def\cross{\times}

\nc{\Sh}{\!\!\!\!/}   \nc{\sh}{\!\!\!/}   
\def\td{\tilde}       
\nc{\half}{\frac{1}{2}}     \nc{\hal}{\tfrac{1}{2}}
  \nc{\thrd}{\frac{1}{3}}     \nc{\thr}{\tfrac{1}{3}}
    \nc{\qrtr}{\frac{1}{4}}     \nc{\qrt}{\tfrac{1}{4}}
\def\one{\bs{1}}

       \def\Im{\trm{Im}}     
    
\def\H{\mcl{H}}

\nc{\Ad}{\ts\tsl{Ad}\ts}
\nc{\ad}{\ts\tsl{ad}\ts}  
\nc{\const}{\trm{const}}  \nc{\spt}{\trm{spt}}

\theoremstyle{definition}                

\begin{document}

\def\Psib{\bar\Psi}  \def\Phib{\bar\Phi}
\def\lamb{\bar\lam}  \def\mub{\bar\mu}

\nc{\dep}[1]{\de_{\eps_{{}_{#1}}}}
\nc{\tdep}[1]{{\tilde\de}_{\eps_{{}_{#1}}}}
\def\gh{{\mit{gh}}}
\def\t{\tilde}

\def\Hcm{\H^{\trm{min}}}
\def\Hm{H^{\trm{min}}}
\def\Im{I^{\trm{min}}}
\def\Imi{I^{\trm{min}}_{\trm{int}}}
\def\G{\mcl{G}}
\def\D{\mrs{D}}
\def\nts{\mspace{-1.7mu}}

\bibliographystyle{perso}
\numberwithin{equation}{section}

\begin{titlepage}

\hfill hep-th/0301222

\vs{3cm}
\bce
\obeylines \tbf{\Large BRST Hamiltonian for Bulk Quantized
Gauge Theory}
\vs{.5cm} Alexander Rutenburg\\
{\em
Courant Institute, New York University,
                  New York,  NY~10002,~USA }\\
{\tt rutenbrg@cims.nyu.edu}
\ece

\begin{abstract}
By treating the bulk--quantized Yang--Mills theory as a constrained
system we obtain a consistent gauge--fixed BRST hamiltonian
in the minimal sector.  This provides an independent
derivation of  the 5--$d$ lagrangian bulk action.
The ground state is independent of the (anti)ghosts
and is interpreted as the solution
of the Fokker--Planck equation, thus establishing a
direct connection to the Fokker--Planck hamiltonian.
The vacuum state correlators are shown to be in agreement
with correlators in lagrangian 5--$d$ formulation.
It is verified  that the complete propagators  remain parabolic
in one--loop dimensional regularization.

\end{abstract}

\vfill\flushright \AmS-\LaTeX{}

\end{titlepage}

\section{\bf{Introduction}}

The usual formulation~\cite{Pesk} of
4--$d$ gauge theory is based on the free
(euclidean) lagrangian action
\bea
S=\qrtr\int_{{\cal M}^4}dx\, F_{\mu\nu}F^{\mu\nu}
\eea
 where
$F_{\mu\nu}=\p_{[\mu} A_{\nu]} + [A_\mu,A_\nu]$ is the curvature
of an $\mrm{SU(N)}$ connection $A$.
In recently introduced
\emph{bulk quantization}~\cite{BaZwQCD,BaZwScal,BaGrZw,BaZwGauge}
(that arose from stochastic quantization approaches
based on ideas of Parisi and Wu~\cite{Stoch,Stoch2})
 one adds an extra fifth nonphysical
dimension $t$ to the spacetime $4$--manifold  ${\cal M}^4$.
All the fields of the theory are then defined on
the extended spacetime
$$\vph(x), \> x \in {\cal M}^4 \>\lgra\>
\vph(t,x), \> (t,x) \in {\mbf R} \cross {\cal M}^4$$
(This $t$ corresponds to the stochastic evolution parameter
or the Monte Carlo iteration time for numerical simulation).
The connection $A_\mu dx^\mu$ is extended to include a fifth
component $A_5dt$ and one has
$F_{5\mu} = \p_5 A_\mu - \p_\mu A_5 +[A_5,A_\mu].$

A set of ghost fields is introduced with two independent
$\mbf Z$--gradings (ghost numbers) $\gh_s$ and $\gh_w$
corresponding to BRST differentials $s$ and $w$, which
raise the respective ghost numbers by one and satisfy
\bea
(s+w)^2=0\>\Ra \>s^2=0,\>w^2=0,\>sw=-ws
\eea
The operator $w$ provides a BRST implementation of the 5--$d$
gauge symmetry, analogous to the usual BRST operator~\cite{Bert,Bau}
(usually called $s$!) connected with
Faddeev--Popov ghosts.  Its cohomology $H^0(w)$ defines
observables. The operator $s$ acts like a rigid
supersymmetry operator and has trivial cohomology.
Observables are \emph{not} required to be $s$--exact.
Fields with
odd total ghost number $\gh \equiv\gh_s +\gh_w$ anticommute.
The action of $s$ and $w$ on the fields is defined as
\begin{align}\label{BrsAlg}
  sA_\mu&=\Psi_\mu & s\Psi_\mu &=0 &
  s\Psib_\mu &=\Pi_\mu & s\Pi_\mu&=0\nn\\
  sA_5&=\Psi_5 & s\Psi_5&=0 &
 s\Psib_5&=\Pi_5 & s\Pi_5&=0\nn\\
   sc&=\Phi & s\Phi&=0 & s\Phib&=\bar c & s\bar c&=0\nn\\
  s\lam&=\mu & s\mu&=0 & s\mub&=\lamb & s\lamb&=0\nn\\[-8pt]
    & & & & & & &
\end{align}\vspace{-34pt}
\begin{align}
 wA_\nu&=D_\nu\lam & w\Psi_\nu = -[&\lam,\Psi_\nu] -D_\nu\mu &
   w\Psib_\nu =& -[\lam,\Psib_\nu]
     & w\Pi_\nu =& -[\lam,\Pi_\nu] + [\mu,\Psib_\nu]\nn\\
wA_5 &=D_5\lam & w\Psi_5 = -[&\lam,\Psi_5] -D_5\mu &
   w\Psib_5 =& -[\lam,\Psib_5]
     & w\Pi_5 =& -[\lam,\Pi_5] + [\mu,\Psib_5]\nn
\end{align}
\vspace{-32pt}
\begin{align}
 w c =& -[\lam,c] -\mu & w\Phi = -[\lam&,\Phi] + [\mu,c] &
   w\Phib =- &[\lam,\Phib] & w\bar c =-&[\lam,\bar c,]
         + [\mu,\Phib]\nn\\
w\lam =& -\hal[\lam,\lam] & w\mu =-[\lam&,\mu] &
   w\mub = -&[\lam,\mub] +\Psib_5 &
   w\lamb = -&[\lam,\lamb] +[\mu,\mub]+\Pi_5\nn
\end{align}

\flush Here $D_\mu = \p_\mu + [A_\mu,\>]$ and
$D_5 = \p_5 + [A_5,\>]$
denotes the usual gauge covariant
derivative.  With some obvious renaming of fields this is
the BRST algebra of~\cite{BaZwGauge}, with a minor \emph{exception}.
To make the action of $w$ on the
quartet $A_5,\Psi_5,\Psib_5,\Pi_5$ symmetric in form,
as it now is, to the action of $w$ on
$A_\mu,\Psi_\mu,\Psib_\mu,\Pi_\mu$ we made the field redefinitions
\beq
\Psib_5 \equiv \bar m + [\lam,\mub] \2\trm{and}\2
\Pi_5 \equiv -l - [\lam,\lamb] + [\mu,\mub].
\eeq
Otherwise one has in~\cite{BaZwGauge}
\begin{align}
  s\bar m&=l & s l &=0 \nn\\
w\bar \lam &= -l & w l &=0 \nn\\
w\mub &= \bar m & w\bar m &= 0\nn
\end{align}
which, although simpler, lacks the aforementioned symmetry and
moreover leads to more cubic ghost interaction terms in $I_{\trm{gf}}$
than our choice here.

The 5--$d$ action for the theory is $s$--exact and $w$--closed
\vspace{-8pt}\bea
w I = 0\vspace{-18pt}
\eea
and is given by
\bea\label{I}
I \!&=&\! I_0 + I_{\mrm{gf}}\nn\\
I_0 \!&\equiv&\! \int d^5x
   s\Big[ \Psib_\mu\big(F^{5\mu} - D_\lam F^{\lam\mu} + \Pi^\mu
    +[\Psib^\mu,c]\big) + \Phib\big(\Psi_5 - a'^{-1}D_\mu\Psi^\mu
-(D_5  - a'^{-1}D^2)c\big)\Big]\nn\\
I_{\mrm{gf}} \!&\equiv&\! \int d^5x\,
      ws\Big[\mub\(A_5 - a^{-1}\p\!\cdot\!A\)\Big]
\eea where $a$ and $a'$ are positive
constant parameters.
After expansion, the $w$--exact
piece $I_{\mrm{gf}}$ fixes the gauge for $A_{\mu}$ and
$\Psi_{\mu}$ to $A_5 = a^{-1}\p\!\cdot\! A$
and $\Psi_5 = a^{-1}\p\!\cdot\! \Psi$.
The theory is well--defined in this gauge
and one has convergence of longitudinal modes.
From the 4--$d$ point of view this
axial type 5--$d$ gauge condition actually
corresponds to an infinitesimal \emph{gauge transformation} $\de
A_\mu = D_\mu a^{-1}\p\cdot A$, so there is no Gribov obstruction
associated with gauge fixing (see~\cite{BaZwGauge}).

Because all free
ghost propagators are retarded, closed ghost loops
vanish (except for tadpoles which can be ignored).  Since
ghost number is conserved, as long as one doesn't compute ghost
correlators the effect of integrating out the ghosts is simply to
suppress the ghosts in the action which, after integrating
out $\Pi_\mu$ as well and rescaling $t$, yields
\beq
I_{\mrm{red}}' = -\qrtr \int d^5x\,\big[
a^{-1}(\p_tA_\mu - D_\mu\p\!\cdot\!A)^2
+ a(D_\lam F^{\lam\mu})^2\big]
\eeq
After analyzing this action in the Landau gauge limit
$a\!\sea\!0$ one finds that the weight is concentrated
in the Gribov region, i.e., where $\p\!\cdot\! A =0$ and
the Faddeev operator is positive,
$-\p\!\cdot\!D(A) \vgeq 0$.  The
physical content of the 4--$d$ theory, such as correlators, is
recovered by going to a time slice $t=\mrm{constant}$.
The reader is referred to~\cite{BaZwGauge} for details.

We will address here the question of finding the proper
hamiltonian corresponding to (\ref{I}).  An outline of how we
proceed is as follows.  We consider just $I_0$, the gauge non--fixed
part of the action, and read off the hamiltonian, which has
simple first class constraints.  One has a choice of whether or
not to include $A_5, \Psi_5, \Psib_5, \Pi_5$ among the canonical
variables; the phase space without these variables is called the
the \emph{minimal sector}. In the hamiltonian formalism the first
class constraints are generators of gauge transformations, and
hence of $w$. To quantize the system one needs a
BRST gauge--fixed hamiltonian. According to homological
BRST theory,
a ghost--antighost pair is introduced for each constraint and used
to construct a BRST generator $\Om$ for $w$, which we choose to do
in the minimal sector for reasons outlined below. We then
obtain a gauge--fixed hamiltonian $\Hm=\Hm_C -\{\Om,K\}$, the gauge
being fixed by the second term with $K$ chosen
so as to give action $I^{\trm{min}}$ (a reduced form of $I$ that
results after integrating out non--minimal fields).

We then go on
to show that the complete ghost propagators remain retarded
in one--loop dimensional regularization.
The retarded character of the full ghost propagators
allows us to establish an equivalence between the
quantum hamiltonian and lagrangian correlation functions.
We also argue that the ground state wave function $P$
has trivial ghost dependence, which provides a direct
connection to the Fokker--Planck equation
\bea
-\int d^4x\, \dd{}{A_\mu(x)}
\Big[\dd{}{A^\mu(x)} -\dd{S_{\mrm{YM}}}{A^\mu(x)}
+a^{-1}D_\mu\p\!\cdot\! A(x)\Big]P(A)=0
\eea

\section{\bf Constrained hamiltonian}
The gauge non--fixed part of the action after expansion is
\bea\label{I0}
I_0    &=& I_F + I_\Pi + I _c \nn\\
        I_F  &=& \int d^5x \big[ \Pi_{\mu}(F^{5\mu}
           - D_{\lam}F^{\lam\mu}) -\bar\Psi_{\mu}\big(
             D^{[5}\Psi^{\mu]}-D_{\lam}D^{[\lam}\Psi^{\mu]}
             -[F^{\mu\nu},\Psi_{\nu}]\big)\big]  \nn\\
I_\Pi  &=& \int d^5x \big[\Pi^2 + 2\Pi_\mu[\Psib^\mu,c]
+ [\bar\Psi^{\mu}, \bar\Psi_{\mu}]\Phi\big] \\
I_c    &=& \int d^5x \,\bar c\big[(\Psi_5 - a'^{-1}D_\mu\Psi^\mu
                  -(D_5  - a'^{-1}D^2)c\Big] \nn\\
        &+& \int d^5x \,\Phib \big[ -(D_5 - a'^{-1}D^2)\Phi
  -[\Psi_5 - a'^{-1}D_\mu\Psi^\mu,c]
  + a'^{-1}[\Psi^\mu,2D_\mu c - \Psi_\mu]\big]\nn
\eea
The gauge fixing term for future reference is
\bml\label{Igf}
I_{\trm{gf}}= \int d^5x\,\big[
              \Pi_5(A^5 - a^{-1}\p\!\cdot\! A)
              + \Psib_5(\Psi^5 - a^{-1}\p_{\mu}\Psi^{\mu})\\
               -\lamb(\p_5 -a^{-1}D\cdot \p)\lam
                -\mub\big( (\p_5-a^{-1}D\cdot\p)\mu
               -a^{-1}\[\Psi^{\mu},\p_{\mu}\lam\]\big)\big]
\end{multline}

First we look at the equations of motion generated
by varying $I_0$ with respect to the fields
$\Psi_5$ and $A_5$
\begin{align}\label{eqm}
0=\dd{I_0}{\Psi_5} &= D_\mu\Psib^\mu + [\Phib,c] -\bar c \equiv \vph_1,
\nn\\[-10pt]{}&{}\\[-10pt]
0=\dd{I_0}{A_5} &= D_\mu\Pi^\mu + [\Psib_\mu,\Psi^\mu] +[\bar c ,c]
+ [\Phib,\Phi] \equiv \vph_2\nn
\end{align}

\flush and obtain what are called \emph{primary constraints}
$\vph_1$ and $\vph_2$.  Note that we use the usual
convention that all functional derivatives with respect
to Grassman fields are \emph{left} derivatives.
It is of interest to observe that the constraints satisfy
\bea s\vph_1 =\vph_2\eea
We substitute these constraints into the action $I_0$, and
obtain a reduced form of the action $I_0^{\trm{min}}$,
 where all terms linear in
$\Psi_5$ and $A_5$ have been eliminated
by the equations of motion (\ref{eqm}).  In this
approach $\Psi_5$ and $A_5$ play the role of lagrange multipliers
and are \emph{not} canonical variables.  This is
analogous to the role $A_0$ plays in enforcing Gauss law $D_j E^j=0$
in the minimal hamiltonian for electromagnetism
or Yang--Mills, where only $A_j$ and $E_j$ are
treated as canonical variables.  Thus
\bea
I_0 =\int d^5x \,\big( \dot A_\mu\Pi^\mu +
                    \dot\Psi_\mu\Psib^\mu + \dot c\bar c
                    - \dot\Phi\Phib - \mcl{H}_C^{\trm{min}}
                    -\Psi^5\vph_1 -A^5\vph_2\big)
\eea
and the reduced form of the action is
\bea
I_0^{\trm{min}} &=& \int d^5x \,\big( \dot A_\mu\Pi^\mu +
                    \dot\Psi_\mu\Psib^\mu + \dot c\bar c
                    - \dot\Phi\Phib - \mcl{H}_C^{\trm{min}} \big),\\
-\mcl{H}_C^{\trm{min}}  &=& \Pi^2  + \big(2[\Psib_\mu,c]
                 -D_\lam F^{\lam\mu}\big)\Pi_\mu
           + \Psib_\mu\big(D_{\lam}D^{[\lam}\Psi^{\mu]}
       +[F^{\mu\nu},\Psi_{\nu}]\big)+[\Psib_\mu,\Psib^\mu]\Phi\nn\\[-5pt]
            & &\\[-10pt]
           & & + a'^{-1} \bar c(D^2 c-D_\mu\Psi^\mu)
    + a'^{-1} \Phib\big(D^2\Phi +[D_\mu\Psi^\mu,c] + [\Psi^\mu,
      2D_\mu c-\Psi_\mu]\big)\nn
\eea

\flush One reads off $H_C^{\trm{min}}$ from $I_0$ by dropping kinetic
terms and setting $\vph_1=\vph_2=0$.  Since $I_0$, as given in
(\ref{I}), is $s$--exact it follows
 that $H_C^{\trm{min}}$ is also
$s$--exact and can be expressed as
\bea\label{HCmin}
H_C^{\trm{min}} = \int d^4x\,
   s\big[ \Psib_\mu\big( \Pi^\mu- D_\lam F^{\lam\mu}
    -[\Psib^\mu,c]\big) -  a'^{-1}\Phib D_\mu\big(
               \Psi^\mu-D^\mu c\big)\big]
\eea

We now proceed with the analysis of this constrained
gauge system which goes according to a standard prescription,
as follows. The reader is referred to to~\cite{Sund,HenTeit}
for background
on constrained systems.  The constrained hamiltonian is written
\beq
\Hm \equiv \Hm_C + \int d^4x\, u^j\vph_j,\2 j=1,2
\eeq
where the $u^j$ are the lagrange
multipliers, here  $\Psi_5$ and $\Pi_5$,
enforcing the primary constraints $\vph_j$.
In general the $u^j$ may be chosen to be included in the
canonical variables.  In our case this would correspond to including the
$A_5,\Psi_5,\Psib_5,\Pi_5$ quartet in the phase space.
Such an approach is termed \emph{nonminimal}.
Our analysis will be based on choosing the more economical
phase space, hence we use the $\Hm$ notation.  One can
in principle consider a nonminimal treatment, but it is
inconvenient for this system and hence remarks related thereto are
relegated to Appendix A.

We use $\Hm_C$ to denote the canonical hamiltonian
with the corresponding action $I_0^{\trm{min}}$.  The action
$I_0$ with constraints is then called the \emph{extended action}
and $\Hm$ is termed \emph{extended hamiltonian}.
The $\approx$ notation is introduced
to represent \emph{weak} equality,
that is equality modulo functions that vanish on the constraint
surface in phase space described by $\vph_j =0$.
$\Hm$ then determines time evolution of all functions $F$
of the fields by
\beq\label{TimeEvol}
\dot F \approx -\{\Hm_C + u^j\vph_j,F\}
\eeq
Here $\{\>,\>\}$ is the graded Poisson bracket.  It is
defined on functions $F$ and $G$ of the fields as
\bml
\{F(x),G(y)\} =
\int d^4z \bigg[ \dd{F(x)}{\vph^a(z)} \dd{G(y)}{p_a(z)}
                 - \dd{F(x)}{p_a(z)} \dd{G(y)}{\vph^a(z)} \\
 + (-)^{\gh(F)}\bigg( \dd{F(x)}{\tha^b(z)} \dd{G(y)}{\pi_b(z)}
         + \dd{F(x)}{\pi_b(z)} \dd{G(y)}{\tha^b(z)} \bigg) \bigg]
\end{multline}
where $\vph^a$ denote all the commuting fields, the $p_a$ their
momenta, and similarly for the odd $\tha^b$ and $\pi_b$.
It satisfies a graded Jacobi identity
\beq
(-)^{\gh(F_{\trm{\tiny 3}})\gh(F_{\trm{\tiny 1}})}
\{ F_1,\{F_2,F_3\}\} + \trm{cyclic perms}=0
\eeq
Note that one has for odd fields
$\{\tha(x),\pi(y)\} = \{\pi(x),\tha(y)\} = -\de(x-y)$.
Moreover, if one expresses $s$ in terms of
functional derivatives
\bea
s=\int d^4x \,\Big[
\Psi^\mu(x)\dd{}{A^\mu(x)}
+\cdots
\Big]
\eea
one finds that $s$ acts as a graded
derivation with respect to the bracket
\bea\label{deriv}
s\{F,G\}=\{s F,G\} + (-)^{\gh(F)}\{F,sG\}
\eea

The constraints $\vph_j$ must be preserved in time, so we
apply (\ref{TimeEvol}) to $\vph_j$ and get
$\{\Hm_C,\vph_j\}\approx 0$ which
generates no further (what would be termed
\emph{secondary}) constraints.
Some computation (the Jacobi identity is useful)
shows that the constraints $\vph_m$ close to generate a
Lie algebra (the structure functions are all constant)
\beq\label{alg1}\bsp
\{ \vph_1^a(x),\vph_2^b(y)\} &= \de(x-y)f^{ab}_{\2 \,c} \vph_1^c(y) \\
\{ \vph_2^a(x),\vph_2^b(y)\} &= \de(x-y)f^{ab}_{\2 \,c} \vph_2^c(y) \\
\end{split}
\eeq
Here the $f^{ab}_{\2 \,c}$ are the (totally antisymmetric)
structure constants of $\mfk{su}(\mrm{N})$.
Note that we expect the second relation to follow from
the first by way of (\ref{deriv}).
It is not hard to check that $\{\Hm_C,\vph_1\}=\{\Hm_C,\vph_2\} =0$.
Thus there are no secondary constraints at all.
Moreover, the constraints are
\emph{irreducible}, meaning that the equations $\vph_m = 0$
are independent.

For constrained hamiltonian systems a functional $F$ whose
bracket with every constraint (including secondary,
if they are present) vanishes weakly
\beq \{\vph_m,F\}\approx 0 \eeq
is said to be \emph{first class}.
First class functionals have important properties.  From
the Jacobi identity it follows that the bracket of first
class functions is first class.  Also, the first class constraints
are the generators of gauge transformations~\cite{HenTeit}
\beq\label{gt}
\delta_\eps F(x) = -\int d^4y\, \eps^m(y)\{\vph_m(y),F(x)\}
\eeq
The ghost numbers of the infinitesimal gauge parameters
$\eps^m$ are chosen so as to leave the ghost number of $F$ invariant,
i.e., $\gh(\eps^m) = -\gh(\vph_m)$.  In our case  $\gh(\eps_1)= 1$
and $\gh(\eps_2)= 0$.
Since all constraints are first class, one need not introduce
Dirac brackets and the analysis of the system is
considerably simplified.
We remark that the analysis of the constraints is in general
highly dependent on where one draws the minimal sector, so
that constraints that are first class in one treatment may
be second class in another, likewise regarding
primary and secondary,
and some constraints may be altogether absent.

Now, for each generator $\vph_m$, the corresponding gauge
transformation is given by (\ref{gt}), but without sum on $m$.
One has
\bal\label{epsalg}
    \dep{2}A_\mu^a &=-(D_\mu\eps_2)^a&
         \dep{1}A_\mu^a &=0\nn\\
    \dep{2}\Psi_\mu^a &=[\eps_2,\Psi_\mu]^a&
         \dep{1}\Psi_\mu^a &=-(D_\mu\eps_1)^a\nn\\
    \dep{2}\Psib_\mu^a &=[\eps_2,\Psib_\mu]^a&
         \dep{1}\Psib_\mu^a &=0\nn\\
    \dep{2}\Pi_\mu^a &=[\eps_2,\Pi_\mu]^a&
         \dep{1}\Pi_\mu^a &=[\eps_1,\Psib_\mu]^a\nn\\
    \dep{2}c^a &=[\eps_2,c]^a&
         \dep{1}c^a &=-\eps_1^a\\
    \dep{2}\Phi^a &=[\eps_2,\Phi]^a&
         \dep{1}\Phi^a &= [\eps_1,c]^a\nn\\
    \dep{2}\Phib^a &=[\eps_2,\Phib]^a&
         \dep{1}\Phib^a&=0\nn\\
    \dep{2}\bar c^a &=[\eps_2,\bar c]^a&
         \dep{1}\bar c^a &= [\eps_1,\Phib]^a\nn
\end{align}
\flush
The full gauge transformations (\ref{gt}) of the fields are given by
\bal
\delta A_\nu&=-D_\nu\eps_2 & \delta\Psi_\nu = [&\eps_2,\Psi_\nu]
-D_\nu\eps_1 &  \delta\Psib_\nu =& [\eps_2,\Psib_\nu]
   & \delta \Pi_\nu =& [\eps_2,\Pi_\nu] + [\eps_1,\Psib_\nu]\nn\\
 \delta c = &[\eps_2,c] -\eps_1 & \delta\Phi = [&\eps_2,\Phi]
+ [\eps_1,c] & \delta\Phib =& [\eps_2,\Phib] &
       \delta\bar c =&[\eps_2,\bar c,]  + [\eps_1,\Phib]
\end{align}
which coincides with the remnant (after the $A_5$ quartet is gone)
of the $w$ algebra (\ref{BrsAlg}) when the
$\eps$ parameters are replaced by variables of opposite statistics
\beq
\eps_3\lgra \mu,\3 \eps_4\lgra -\lam,\3 s\lam = \mu
\eeq
with $\gh_s(\lam)=0$, $\gh_w(\lam) = 1$ and
$\gh_w(\mu) = \gh_s(\mu) = 1$.
In~\cite{BaZwGauge} transformation properties
were \emph{imposed} on the fields by hand, as each new field was
added to the action, in such a way as to have $w$ invariance of
the action $I$.
From our point of view the gauge algebra and field transformations
are in fact \emph{determined} by the constraints, i.e., the action.

As for the lagrange multiplier fields, one
has the freedom of assigning to them any gauge transformation
properties one sees fit, and we may therefore choose to
transform them in such a way as to make the entire action
\bea
I_0=I_0^{\mrm{min}}-\int d^5x\,(\Psi_5 \vph_1 +A_5 \vph_2)
\eea
gauge invariant.  This can always be arranged
even in the most general cases with second class
constraints~\cite{HenTeit}, and in our case amounts to
(not surprisingly) setting
\bea
\de\Psi_5 = -D_5 \eps_1 +[\eps_2,\Psi_5] ,\3
\de A_5 = -D_5 \eps_2
\eea

\section{ \bf Minimal BRST hamiltonian}

The following considerations are direct consequences of standard
results of homological BRST theory (we refer to~\cite{HenTeit} for
details). The extended phase space is introduced by including in the
minimal sector a ghost--antighost conjugate
pair for each of the constraints
$\vph_1$ and $\vph_2$. Hence we add $\mu,\mub$ for $\vph_1$ and
$\lam,\lamb$ for $\vph_2$, and as the notation indicates, identify
them with the ghost fields in $I_{\trm{gf}}$. The
corresponding kinetic terms $-\dot\mu\mub + \dot\lam\lamb$ are
included in the action. Note the minus
sign in $\{\mu(x),\mub(y)\}= -\de(x-y)$.

By inspection of (\ref{BrsAlg}), one easily finds the
generator $Q$ for $s$ on the extended phase space
\bea\label{Q}
s=-\{Q, \>\}, \3  Q = \int d^4x\,\big(\Psi_\mu\Pi^\mu +
                     \Phi\bar c +\mu\lamb\big)
\eea
Therefore we have
\beq
H_C^{\trm{min}} = -\{Q,X\},\3
    X =-\int d^4x\,\big[\Psib_\mu\big( \Pi^\mu- D_\lam F^{\lam\mu}
    +[\Psib^\mu,c]\big) -  a'^{-1}\Phib D_\mu\big(
               \Psi^\mu-D^\mu c\big)\big]
\eeq
A main theorem of BRST theory~\cite{HenTeit}
provides the existence of a
BRST generator $\Om$ for $w$ which, because  the $\vph_j$
generate a genuine Lie algebra, takes a particularly simple form
\bea\label{Om}
w = -\{\Om, \>\},\3 \Om \!\!&=&\!\! \int d^4x\,\big(\mu\vph_1
  - \lam \vph_2 -  \hal\lamb[\lam,\lam]
                  -\mub[\mu,\lam]\big)\nn\\[-5pt]
            & &\\[-10pt]
            \!\!&=&\!\! -\{Q,\int d^4x\,\big(\lam\vph_1
-\hal \mub[\lam,\lam]\big)\}\nn
\eea

\flush Thus $\Om$ is $Q$--exact which implies
$\{Q,\Om\}=0$, as expected.
We remark that in theories where the constraints do not
generate a closed algebra with constant structure functions,
the BRST generator may be much more complicated (an infinite
series expansion in a ghost degree).  So again, we see the
attractive simplicity of this 5--$d$ theory.

One needs to construct an appropriate BRST invariant
extension of $\Hm_C$ and then gauge--fix it.  But $\Hm_C$ is already
$w$--invariant so the gauge--fixed
BRST hamiltonian corresponding to $I=I_0 + I_{\trm{gf}}$, as
expressed in terms of this minimal set of fields, is
then simply given by
\bea
H^{\trm{min}} = H_C^{\trm{min}} + H^{\trm{min}}_{\trm{gf}}
= -\{Q,X\} - \{\Om, K\}
\eea
where the \emph{gauge fixing fermion} $K$, as it
is frequently called, is
chosen to be
\beq
K = -\{Q,f\}\equiv -\{Q,a^{-1}\!\!\int d^4x\,\mub\p\!\cdot \!A\} =
a^{-1}\!\!\int d^4x\,\big(\lamb\p\!\cdot\! A
+ \mub\p\!\cdot\!\Psi\big)
\eeq
From the Jacobi identity and $\{\Om,Q\}=0$ one has
\beq
\{\Om,K\} = \{Q,\{\Om,f\}\}
\eeq
so $H^{\trm{min}}$ can be written in the compact form
\beq
H^{\trm{min}} = -\{Q,\bar Q\},\3\bar Q =X + \{\Om,f\}
\eeq

Explicitly one has
\bea
-\{\Om,K\} &=& a^{-1}\!\!\int d^4x\,\big(\vph_1\p\!\cdot\!\Psi
- \vph_2\p\!\cdot\!A -\mub D\!\cdot\!\p\mu-\lamb D\!\cdot\!\p\lam
-\mub[\Psib^\mu,\p_\mu\lam]\big)
\eea
Thus the BRST hamiltonian in its fully expanded form is
\bea
H^{\trm{min}} &=&-\{Q,\bar Q\}\nn\\
             &=&-\{Q,X\} - \{\Om, K\}\nn\\
          &=&\int d^4x\,-\Big(\Pi^2  + \Pi_\mu\big(2[\Psib_\mu,c]
                 -D_\lam F^{\lam\mu}\big)
           + \Psib_\mu\big(D_{\lam}D^{[\lam}\Psi^{\mu]}
         +[F^{\mu\nu},\Psi_{\nu}]\big)+[\Psib_\mu,\Psib^\mu]\Phi\nn\\
           & & \qquad+ \frac{1}{a'} \big[\bar c(D^2 c-D_\mu\Psi^\mu)
    +  \Phib\big(D^2\Phi +[D_\mu\Psi^\mu,c] + [\Psi^\mu,
      2D_\mu c-\Psi_\mu]\big)\big]\Big)\\
     & & \qquad+\frac{1}{a}\Big((D_\mu\Psib^\mu
              + [\Phib,c] -\bar c)\p\!\cdot\!\Psi
              - (D_\mu\Pi^\mu + [\Psib_\mu,\Psi^\mu] +[\bar c ,c]
              + [\Phib,\Phi] )\p\!\cdot\!A\nn\\
     & & \bspc\spc-\mub D\!\cdot\!\p\mu-\lamb D\!\cdot\!\p\lam
      -\mub[\Psib^\mu,\p_\mu\lam]\Big)\nn
\eea
Our canonical treatment agrees with~\cite{BaZwGauge}
because one can easily check that after
integrating out $\Pi_5, A_5, \Psib_5, \Psi_5$ in the
lagrangian action $I$ [= (\ref{I0}) + (\ref{Igf})] one
gets precisely
\bea
I^{\trm{min}} = \int d^5x\,\big(
      \dot A_\mu\Pi^\mu + \dot\Psi_\mu\Psib^\mu
      + \dot c\bar c - \dot\Phi\Phib +\dot\lam\lamb -\dot\mu\mub
            -\mcl{H}^{\trm{min}}\big)
\eea
So in fact, what we have done here is give a consistent constructive
derivation of the (reduced form of) action $I$ based on the
canonical analysis of the constrained hamiltonian.

\section{\bf Propagators}
In this section we study the propagators
and show that all complete ghost (and ghost of ghost)
propagators stay retarded in one-loop
dimensional regularization.  This is an important feature
of bulk quantization and will be key to establishing
the advertised results on correlators ant the
ground state in the next section..

Let us then begin by first computing the \emph{free}
propagators
by inverting the quadratic part of the action $I^{\trm{min}}$,
which is given by the quadratic form
\bml\label{Im2}
\Im_{(0)} = -\int d^5x\,\, A^\mu(-\de_{\mu\nu}\p_5
+ a^{-1}\p_\mu\p_\nu +\bx_{\mu\nu}^{\trm{tr}})\Pi^\nu -\Pi^2\\
+ \Phib(\p_5-{a'}^{-1}\bx)\Phi + \lamb(\p_5-{a'}^{-1}\bx)\lam
+ \mub(\p_5-{a'}^{-1}\bx)\mu\\
+ \bs{\big(}\Psib_\mu,\bar c\bs{\big)}
\bpmx
\de_{\mu\nu}\p_5 - a^{-1}\p_\mu\p_\nu -\bx_{\mu\nu}^{\trm{tr}} & 0\\
(a-a')\p_\nu/aa'            &  \p_5-{a'}^{-1}\bx
\epmx
\bpmx \Psi_\nu \\ c \epmx
\end{multline}
Here
\bea
\bx_{\mu\nu}^{\trm{tr}}=\bx P^{\mrm{tr}}
=\de_{\mu\nu}\bx-\p_\mu\p_\nu
\eea
and $P^{\mrm{tr}}$ and $P^{\mrm{lg}}$ are
the usual transverse and longitudinal projectors.
They may be defined via their Fourier transforms
(denoted by $\hat{\>\,}$)
\bea
\hat P^{\mrm{tr}}_{\mu\nu}(p)= \delta_{\mu\nu}
- \frac{p_\mu p_\nu}{p^2},\qquad \hat P^{\mrm{lg}}_{\mu\nu}(p)
= \hat P^{\mrm{tr}\perp}_{\mu\nu}(p) = \frac{p_\mu p_\nu}{p^2}.
\eea
and provide an orthogonal decomposition
$\mbf{1}=P^{\mrm{tr}}(\p)+ P^{\mrm{lg}}(\p).$

The single blocks are trivial to invert and give the free
momentum space propagators
\bea
\hat {\mcl{D}}_{0,\Phi\Phib}(p)=\frac{1}{ip_5+p^2/a'},\qquad
\hat {\mcl{D}}_{0,\lam\lamb}(p)=\hat {\mcl{D}}_{0,\mu\mub}(p)=
\frac{1}{ip_5+p^2/a}
\eea
The $2\cross 2$ block is also straightforward to invert
(after using integration by parts to generate a lower
left term).  The
non-vanishing transverse free propagators are then
\bea
\hat {\mcl{D}}_{0,A^\mu A^\nu}^{\mrm{tr}}(p)=
\frac{2P^{\trm{tr}}_{\mu\nu}}{p_5^2+(p^2)^2},\qquad
\hat {\mcl{D}}_{0,A^\mu\Pi^\nu}^{\mrm{tr}}(p)=
-\hat {\mcl{D}}_{0,\Pi^\mu A^\nu}^{\mrm{tr}}(p)=
\frac{2P^{\trm{tr}}_{\mu\nu}}{ip_5+p^2}\2
\eea
Using the Hodge decomposition for vector fields
\beq
X_{\mu} = X_{\mu}^{\mrm{tr}} + \p_{\mu}F
\eeq
the longitudinal piece is easily inverted as well.
The result is
\bea
\hat {\mcl{D}}_{0,A^\mu A^\nu}^{\mrm{lg}}(p)=
\frac{2P^{\trm{lg}}_{\mu\nu}}{p_5^2+(p^2)^2/a^2},\2
\hat {\mcl{D}}_{0,A^\mu\Pi^\nu}^{\mrm{lg}}(p)=
-\hat {\mcl{D}}_{0,\Pi^\mu A^\nu}^{\mrm{lg}}(p)=
\frac{2P^{\trm{lg}}_{\mu\nu}}{ip_5+p^2/a},\2
\hat {\mcl{D}}_{0,\Pi^\mu\Pi^\nu}=0
\eea
Note that if $p_5$ is integrated  out in
$\hat {\mcl{D}}_{0,A^\mu A^\nu}$,
the correct 4--$d$ Yang--Mills propagator is recovered.
The $(\Psi,c)$ $2\cross 2$ block similarly gives the following
non--vanishing free propagators
\bal
\hat {\mcl{D}}_{0,\Psi^\mu \Psib^\nu}^{\mrm{tr}}(p)&=
\frac{\de_{\mu\nu}}{ip_5+p^2}&
\hat {\mcl{D}}_{0,\Psi^\mu\Psib^\nu}^{\mrm{lg}}(p)&=
\frac{\de_{\mu\nu}}{ip_5+p^2/a}\\
\hat {\mcl{D}}_{0,c\Psib^\nu}^{\mrm{lg}}(p)&=\frac{a'-a}{aa'}
\frac{p_\nu}{(ip_5+p^2/a)(ip_5+p^2/a')}&
\hat {\mcl{D}}_{0,c\bar c}(p) &=\frac{1}{ip_5+p^2/a'}
\end{align}

Upon taking the inverse Fourier transform
\bea
{\mcl{D}}_0(t,x) = \frac{1}{(2\pi)^5}\int\! dp_5\, e^{itp_5}
\!\int\!d^4p\,e^{ix\cdot p} \hat {\mcl{D}}_0(p_5,p)
\eea
one sees
that all the \emph{free} propagators for the
ghosts (and ghosts of ghosts)
are retarded, since there is no pole in the upper $p_5$ half--plane
($a$ and $a'$ are positive) and closing the $p_5$ contour in the
lower half--plane gives $\tha(t)$.

What about the complete propagators $\mcl{D}$ then?
For motivation consider the
Green's functions $G_0$ and $G$ (for $\lam,\lamb$ say), satisfying
\bea
\bay{cc}
\>(\p_t - a^{-1}\p^\mu \p_\mu)G_0(t-s;x-y) \<&=\,\de(s,y)\\
(\p_t - a^{-1}D^\mu \p_\mu)G(t-s;x,y;A) \<&=\,\de(s,y)
\eay
\eea
Of course $G_0=\mcl{D}_0$ is just the free propagator, but
$G\neq \mcl{D}=\bra T\lam(t,x)\lamb(s,y)\ket_{I^{\mrm{min}}}$ since
the (time ordered)
correlator involves integration over $\D\nts A$ as well.
Nevertheless,
it is instructive to look at properties of $G$
prior to integration.
From Duhamel's principle~\cite{PDE1} we have the
following convolution
relation between $G$ and $G_0$.
\begin{multline}
G(t-s;x,y;A) = G_0(t-s;x-y)
+ \int_s^t d\tau\int d\xi\,
G_0(t-\tau;x-\xi) [A^\mu(\xi),\p_\mu G(\tau;\xi,A)]
\end{multline}
where we have suppressed indices. From this we
see that not only is $G$ automatically retarded as well, but for a
sufficiently regular $A$ one would conclude that
\bea
\lim_{t\sea s} G(t-s;x,y;A) = \lim_{t\sea s} G_0(t-s;x-y) =
\de(x-y)
\eea
The significance is that canonical commutation relations
are \emph{formally} satisfied if one assumes
regularity of $A$ and
\bea
\lim_{t\sea s}\mcl{D}(t-s;x,y)\equiv
\lim_{t\sea s}\bra G(t-s;x,y;A)\ket
=\bra \lim_{t\sea s}G(t-s;x,y;A)\ket
\eea
However, one cannot assume this so a separate argument is required
to check the divergent case.
We now proceed to show that the conclusion indeed applies
to $\mcl{D}$ as well, for $d < 4$ by dimensional regularization
at one--loop level. This amounts to evaluating diagrams like this one,
which is the first-order correction to the free $\Psi\Psib$ propagator
\\[20pt]
\bea
\<\<\<\<\<\<\<\<\<\<\hat\Sig(E,p)&=&\\[-47pt]
& & \>{\begin{fmffile}{sigma}
\begin{fmfgraph*}(150,75)
\fmfpen{thin}
\fmfleft{i1}
\fmfright{o1}
\fmf{plain,label=$E,,p$}{i1,v1}
 \fmf{plain}{v1,v2}
 \fmf{plain}{v2,v3}
  \fmf{plain,right=0,label=$E\!+\!\om,,p\!+\!k$}{v3,v4}
  \fmf{plain,right=0}{v4,v5}
  \fmf{plain}{v5,v6}
  \fmf{plain,right=0,label=$E,,p$}{v6,o1}
  \fmffreeze
  \fmf{boson,left=.7,label=$\omega,,k$}{v2,v5}
\end{fmfgraph*}
\end{fmffile}}\nn\\
&=& \frac{1}{a^2}\int\frac{d\om\,d^dk}{(2\pi)^{d+1}}\,
p^\mu\bigg[\frac{2}{\om^2+(k^2)^2}
              \hat P_{\mu\nu}^{\trm{tr}}(k)\nn\\[-5pt]
& & \\[-5pt]
& & \spc\qquad\quad +\frac{2}{\om^2+(k^2)^2/a^2}
\hat P_{\mu\nu}^{\trm{lg}}(k)\bigg]
\frac{(p+k)^\nu}{(i E+\om)+(p+k)^2/a}\nn
\eea

\flush with
\bea
\hat {\mcl{D}} =\hat {\mcl{D}}_0 +
\hat {\mcl{D}}_0\hat\Sig\hat {\mcl{D}}_0 +\cdots
=\hat {\mcl{D}}_0+\hat {\mcl{D}}_1 +\cdots
\eea
  Let us
outline the computation of $\hat {\mcl{D}}_1^{\trm{tr}}(E,p)$,
the evaluation of the longitudinal component being similar.
Since this is a standard but lengthy diagram integral regularization,
for reasons of continuity we omit full detail here,
which may be found in Appendix B.

Expressing the denominators as parameter integrals
\beq\label{ParInt}
\frac{1}{B^z} =\frac{1}{\Gam(z)}
\int_0^\infty d\al\, \al^{z-1}e^{-B\al}
\eeq
one obtains after some computation the following expression
\begin{multline}
\hat{\mcl{D}}^{\trm{tr}}_1(t,p)
= \Gam(\eps)\frac{(d\!-\!1)p^2}{d(16\pi)^{d/2}}
\int_0^\frac{1}{1+a}\frac{d\al}{\al^\eps}\,[1\!-\!(1\!+\!a)\al]
\qquad\qquad\\
\int\frac{dE}{2\pi}\,\frac{e^{iEt}}
              {[(1\!-\!\al)p^2+iEa]^\eps(iE+p^2/a)^2}
\end{multline}
where $2\eps\equiv 4-d$ and we have taken the inverse Fourier
transform in $E$.
Using parameter integrals again and integrating gives
\bea
\frac{d\!-\!1}{(1\!+\!a)d(16\pi)^{d/2}}
     \Big(\frac{1\!+\!a}{a}\,t\Big)^\eps
         e^{-p^2t/a}\big(p^2t/a\big)
\int_0^1\frac{d\al}{\al^\eps}\,(1\!-\!\al)
\int_0^1\frac{dx}{x} \,x^\eps(1\!-\!x) e^{-\al\frac{p^2t}{a(1+a)}x}
\eea
$\Gam(\eps)$ being cancelled by writing
$[(1\!-\!\al)p^2+iEa]^\eps$ as a parameter integral. Expanding, we find
\bea
\hat{\mcl{D}}^{\trm{tr}}_1(t,p) = \frac{3/2}{(32\pi)^2}
e^{-p^2t/a}\frac{p^2t}{a(1\!+\!a)}\Big[
\frac{1}{\eps}+1-\log\frac{t}{a(1\!+\!a)} +O(\eps\log t)\Big]
\eea
and so $\lim_{t\sea 0}\hat{\mcl{D}}^{\trm{tr}}_1(t,p)=0$.

Since (apart from
the gauge parameters $a$ and $a'$) all the ghost propagators
except $\hat {\mcl{D}}_{0,c\Psib}$ are of the same
form it is clear that these considerations also apply to them.  And
if one rewrites ${\mcl{D}}_{0,c\Psib^\nu}^{\text{lg}}$ as
\bea
\frac{p_\nu}{p^2}\Big(\frac{1}{ip_5+p^2/a'}
-\frac{1}{ip_5+p^2/a}\Big)
\eea
it is not difficult to see that the result will be true here as well.
Thus our conclusion applies to all ghost propagators.

\section{Ground State}

We now turn to the \emph{ground state} $P$ of the theory.
It is the zero eigenvector of the hamiltonian
\beq
\hat HP=0
\eeq
normalized to be a probability density, $\disp\int \D\nts M\,P=1$,
where $\D\nts M$ is a functional measure determined
below and $\hat H$ is an appropriate operator form of
$H^{\trm{min}}$.
In this section we will show that $P$ is just the ground state
of the Fokker--Planck hamiltonian.
Also, by showing that
the equal time limit of correlators of the 5--$d$ theory agrees
with the corresponding expectation value with respect to $P$,
we will use this to
establish equivalence between the hamiltonian
and bulk lagrangian quantization.

We take the operator representation
of the fields to be
\bea \Pi^\mu(x) = \dd{}{A_\mu(x)},\3\Psi^\mu(x)=\dd{}{\Psib_\mu(x)},
\3\Phi(x) = \dd{}{\Phib(x)},\3\text{etc.}
\eea
with all \emph{ghosts} acting as functional derivatives with respect
to the corresponding \emph{antighosts} (similarly for ghosts of ghosts).
Due to signs, we must choose $\Phi(x)=-\de/\de\Phib(x)$
and $\mu(x)=-\de/\de\mub(x)$. This achieves the correspondence
between the graded operator commutator and graded bracket via
\vs{-8pt}
\bea\LL\>,\>\RR=-\{\>\,,\>\}\eea
Recall that $[\xi,\eta]^a=f^a_{\>\,bc}\xi^b\eta^c$
still denotes the Lie algebra (not operator) commutator.

In this representation correlators of operators $\mcl{O}$ are
computed by integrating over $\D\nts M$ with weight $P$, where
$\D\nts M$ is the measure obtained after integrating out all ghosts in
\bea\int\,\D\nts A \,\D\Pi\,\D\Phi \D\Phib\,\D\nts\mu \D\nts\mub
\,\D\Psi \D\Psib\,\D c\D\bar{c}\,\D\nts\lam \D\nts\lamb\,\exp(\Im)
\eea
That the result is of the form
\bea\label{DM}
\bay{rl}
\D\nts M \!\!& =\, \D\nts A\Sig(A)\,
\D\Psib\Psib\, \D\nts \lamb\lamb\,
\D\bar c\bar c\, \D\Phib\de(\Phib)\,\D\mub\de(\mub)\\[5pt]
 & = \,\D\nts A \Sig(A)\,
\D\Psib\de(\Psib)\, \D\nts \lamb\de(\lamb)\,
\D\bar c\de(\bar c)\, \D\Phib\de(\Phib)\,\D\nts \mub\de(\mub)
\eay
\eea
can be seen as follows. (Note that the second equality
is simply the fact that in fermionic calculus
$d\tha\,\tha=d\tha\,\de(\tha)$.)
First, integrate out $\mu$.  This
contributes
\bea
{\det}^{-1}\Big(\p_5-D\!\cdot\!\p\Big) \D\nts \mub \de(\mub)
\eea
and therefore
kills the cubic term $\disp\frac{1}{a}\mub[\Psi^\nu,\p_\nu\lam]$.
Then integrate out $\lam$ which contributes
\bea
\det\Big(\p_5-D\!\cdot\!\p\Big)\D\nts\lamb\de(\lamb)
\eea
The determinants cancel, the $\lam,\lamb,\mu,\mub$ quartet
dependence is gone from $\Im$, and the contribution
to $\D\nts M$ is
\bea
\D\nts\lamb\lamb\,\D\nts\mub\,\nts\de(\mub)
\eea

Now for the rest of the measure, as $\Im$ is quadratic
in $\Pi$, one can integrate $\Pi$ out to obtain
\bea
\int \D\nts A\,\D\Phi \D\Phib\,\D\Psi \D\Psib\,\D c\D\bar{c}\,
\exp\big({\Im_{(0)}}'+{\Imi}'+I_{\trm{red}}
\big)
\eea
where
\bea
I_{\trm{red}}
=-\int d^5x\,\big(\p_tA_\mu-\frac{1}{a}D_\mu\p\!\cdot\!A
-D^\lam F_{\lam\mu}\big)^2
\eea
and similarly to (\ref{Im2}) one has a quadratic form
\bea
{\Im_{(0)}}'= -\int d^5x\,\big(\Psib^\mu,\bar c,\Phib\big)
     \bpmx \mcl{A}_{\mu\nu} & \mcl{B}_\mu & 0\\
           \mcl{C}_\nu      &   \mcl{D}   & 0\\
                   0        &     0       & \mcl{D}  \epmx
     \bpmx \Psi^\nu\\ c \\ \Phi\epmx
\eea
but with the full $A$--dependence kept in the operators
\bea
\bay{ccl}
\mcl{A}_{\mu\nu} \!&=& \!\!\disp{\de_{\mu\nu}\p_5-\frac{1}{a}D_\mu\p_\nu
          +\frac{1}{a}\de_{\mu\nu} -[F_{\mu\nu},\,\bcdot\,]
                       -\de_{\mu\nu}D^2+D_\nu D_\mu}\\
\mcl{B}_{\mu} \<&=& \!\!\disp{2[\dot A_\mu
               -D^\lam F_{\lam\mu},\,\bcdot\,]}\\
\mcl{C}_{\nu} \<&=& \!\!\disp{\frac{1}{a}\p_\nu-\frac{1}{a'}D_\nu}\\
\mcl{D}       \<&=& \!\!\disp{\p_5 -\frac{1}{a'}D^2
                  +\frac{1}{a}[\p\!\cdot\!A,\,\bcdot\,]}
\eay
\eea
and the cubic and quartic ghost interactions are collected in
\bml
{\Imi}' =\int d^5x\,\Big(
-\frac{1}{a}\Phib[\p\!\cdot\!\Psi,c]
+\frac{1}{a'}\Phib[D_\mu\Psi^\mu,c]
-\frac{1}{a'}\Phib[\Psi^\mu,\Psi_\mu]\\
+[\Psib_\mu,\Psib^\mu]\Phi
+\frac{2}{a}\Phib[\Psi^\mu,D_\mu c]
-[\Psib^\mu,c][\Psib_\mu,c]\Big)
\end{multline}

Now observe that
\bea
\bpmx \mcl{A}_{\mu\nu} & \mcl{B}_\mu & 0\\
           \mcl{C}_\nu      &   \mcl{D}   & 0\\
                   0        &     0       & \mcl{D}  \epmx
=\pp{}{t}-L_0(\p)+L_{\trm{int}}(A)
\eea
is a perturbation of the parabolic operator
$\p/\p t-L_0$ by $L_{\trm{int}}(A)$.
Let's ignore the ghost interactions ${\Imi}'$ for the moment.
Since $G_0\equiv(\p/\p t-L_0
+L_{\trm{int}}(A))^{-1}$
is retarded one can use the arguments of~\cite{BaZwScal}
to expand
\bea
\bay {rl}
\det\Big(\disp\pp{}{t}-L_0+L_{\trm{int}}(A)\Big)
&\!=\>\const\cdot\exp\mrm{Tr}\log(\one +G_0L_{\trm{int}}(A))\\
&\!=\>\exp\mrm{Tr}\big(L_{\trm{int}}G_0
      -\disp\half L_{\trm{int}}G_0L_{\trm{int}}G_0+\cdots\big)
\eay\eea
and only the $A$--dependent `tadpole'
$\exp\mrm{Tr}(L_{\trm{int}}G_0)$ survives.
One sees that integrating out $\Psi,c,\Phi$ will generate
just delta functions of $\Psib,\bar c, \Phib$ times
the $A$--dependent tadpole term.
Including the ghost interactions into $L_{\trm{int}}$ gives a
ghost--dependent tadpole, which we ignore.

The upshot is that indeed, after including the
$\lamb,\mub$ component,
the measure must be of form (\ref{DM}), where
$\Sig(A)$ involves ${\Im_{(0)}}'$ (and tadpoles).
This is not surprising for a measure that ought to
give non-trivial results for quantities with zero ghost number.
We shall not attempt to investigate $\Sig(A)$ in detail, but
we expect that after restriction to a time slice it
will generate $P(A)$.

Since $\gh(\hat H)=\gh(\Hm)=0$ the hamiltonian preserves ghost
number. (The actual form, i.e., operator ordering, of $\hat H$
will be dealt with below and is irrelevant for now.)
 Thus, when $P$ is expanded in ghost degree
\def\oset{\overset}
\bea
P=\sum_j \oset{(j)}{P},\3\gh(\oset{(j)}{P})=j
\eea
each component solves the equation separately
\bea
\hat H \oset{(j)}{P}=0
\eea
But because the measure $\D\nts M$ kills terms with non--zero
ghost number
one has
\vspace{-5pt}
\bea
\int \D\nts M\,P=\int \D\nts M\,\oset{(0)}{P}
\\[-16pt]\nn\eea
so only $\oset{(0)}{P}$ contributes and we impose the
condition that the ground state is \emph{independent of the antighosts}
\bea
P=\oset{(0)}{P}=\oset{(0)}{P}(A)
\eea
in agreement with the remark above regarding $\Sig$ and $P$.
Also, since $P$ only depends on $A$ it follows that it must
satisfy
\bea
\hat QP=\hat \Om P=0
\eea
where $\hat Q$ and $\hat \Om$ are obtained from (\ref{Q}) and
(\ref{Om}) by replacing all ghosts and $\Pi$
by derivatives.

We now come to operator ordering.  Propagators are time ordered
correlators, so for any fields $\vPsi_j$ one has
\bea
\bra T\vPsi_1(x)\vPsi_2(y)\ket = \tha(x_5-y_5)\bra\vPsi_1(x)
\vPsi_2(y)\ket + (-)^{\gh(\vPsi_1)\gh(\vPsi_2)}\tha(y_5-x_5)
\bra\vPsi_2(y)\vPsi_1(x)\ket
\eea
But we have determined that the full ghost
propagators are retarded, i.e., proportional to the $\tha$ function,
so one has for the $\lam$--$\lamb$ propagator, for example
\bea
\bay{cc}
\tha(x_5-y_5)\sim \bra T\lam(x)\lamb(y)\ket
= \tha(x_5-y_5)\bra \lam(x)\lamb(y)\ket
-\tha(y_5-x_5)\bra\lamb(y)\lam(x)\ket\\
\Ra\2\bra\lamb(y)\lam(x)\ket =0
\eay
\eea
and similarly for the other ghosts (and ghosts of ghosts).
One then has agreement with the hamiltonian representation
\bea
\lim_{\delta t\ra 0}\bra \lam(t +\delta t,x)
\lamb(t,y)\ket
=\de(x-y)=\int \D\nts A\cdots \D\nts\lamb\lamb
\dd{}{\lamb(x)}\lamb(y)P(A)
\eea
Similarly for the $\mu$--$\mub$ correlator one has
\bea
\lim_{\delta t\ra 0}\bra \mu(t +\delta t,x)
\mub(t,y)\ket
=\de(x-y)=\int \D\nts A\cdots \D\nts\mub\de(\mub)
\dd{}{\mub(x)}\mub(y)P(A)
\eea
and so on.  So the effective hamiltonian ordering prescription is
the time ordering.
All \emph{fields} go to the \emph{right}, that is
all derivative operators to the left.
One readily checks that this
is consistent with the $\Pi$--$A$ propagator as well.

Finally, noting that $\Pi(x)=\displaystyle\dd{}{A(x)}$ is
ordered to the left in $\hat H$ we find that $P(A)$ solves
the Fokker--Planck equation
\bea
\hat H_{\trm{FP}}(A)P(A) \equiv -\int d^4x\, \dd{}{A^\mu(x)}
\Big[\dd{}{A_\mu(x)} -K^\mu(x;A)\Big]P(A)=0,\\
K^\mu(x;A)\equiv D_\lam F^{\lam\mu}(x) +a^{-1}D^\mu\p\!\cdot\!A(x)
=\dd{S_{\mrm{YM}}}{A_\mu(x)} +a^{-1}D^\mu\p\!\cdot\! A(x)
\eea
since $\hat H$ is effectively equal to precisely
$\hat H_{\trm{FP}}$ when
acting on functionals of $A$ only.  Of course the above considerations
do not preclude degeneracy, i.e., we have not proven uniqueness
here.


\section{Conclusion}
We treated the bulk--quantized gauge theory
as a constrained gauge system and
found that the canonical analysis of what happen to be particularly
simple constraints leads directly to a BRST gauge--fixed hamiltonian
and a corresponding action that agrees with (a reduced form of)
the bulk action~\cite{BaZwGauge} arrived at in the lagrangian
formulation.  The hamiltonian is $s$--exact and $w$--closed
\bea
H^{\trm{min}} = -\{Q,\bar Q\}= -\{Q,X+\Om f\}
\eea
The lagrange multiplier fields for fixing the gauge were not
included among the canonical variables
(which we consider inconvenient
due to a larger gauge algebra), but we made some
elementary observations about how one may in principle
proceed with inclusion of the lagrange multipliers
in the phase space.

By dimensionally regularizing the
self--energy one--loop correction to a
representative ghost propagator we have concluded that
the complete propagators for all ghosts (and ghosts of ghosts)
are indeed retarded.
Consequently we found that the ground state $P$ depends on $A$ only
and is in fact just the
ground state of the Fokker--Planck hamiltonian $P=P_{\trm{FP}}$.
We have also displayed the consistency of the hamiltonian
formulation at the quantum level in that expectation
values with respect to the ground state $P$ are compatible with the
expectation values with respect to the 5--$d$ action $I$.

Interesting questions to consider in the future may be
renormalization of the equal--time theory governed by the
Fokker--Planck equation and relation thereof to the
renormalization of the 5--$d$ theory governed by $I$.

\section{Acknowledgements}
The author is very grateful to Daniel Zwanziger for his
invaluable help and support, without which this article
would not have been possible.  Many thanks to Laurent Baulieu
for illuminating discussions.

\section{Appendix A: Nonminimal approach}

In our opinion, the nonminimal treatment, which is based on a
larger action
\bea I_0^E = I_0 -\sum_{j=1}^4u^j\td \vph_j \eea
with more constraints and a larger phase space,
does not lend itself to convenient quantization for the following
reasons.
As we shall see below, the gauge algebra contains 4 independent
$\eps$ parameters (which we may reduce to 2 by hand).
Therefore the BRST implementation of the
nonminimal gauge symmetry would necessarily
involve 2 extra conjugate ghost pairs
(in addition to $\lam,\lamb$ and
$\mu,\mub$).  Then one would need to find a proper gauge fixing
fermion $K$ that would give the same action as $I$ (after
integrating out the extra fields).  Given the already
imposing field content of the theory one would want to avoid
bringing in more fields. In addition,
the identification of all the ghosts associated with constraints
with the ghosts in the lagrangian action $I_{\trm{gf}}$
may become tenuous since such
identifications depend on the particular
gauge fixing.  For the minimal case the situation was
quite simple regarding these issues, hence our choice.
Nevertheless, in case there is further interest in the
nonminimal direction, for completeness we include here
a brief discussion of how one may approach the nonminimal
treatment.

Since $I_0$ is already in first order form we can write
\bea I_0 = \int d^5x \big( \dot A_\mu\Pi^\mu + \dot\Psi_\mu\Psib^\mu
        + \dot c\bar c - \dot\Phi\Phib - \mcl{H}_C \big)
\eea where $\cal{H}_C$ can be written out explicitly, but we don't
need it now. We can immediately read off the canonical momenta
\bea \Pi_{A_\mu}\equiv  \frac{\de I_0}{\de \dot A_\mu} =\Pi^\mu,
\3 \Pi_{\Psi_\mu}\equiv\frac{\de I_0}{\de\dot\Psi_\mu} =\Psib^\mu,
\3 \Pi_{c}\equiv\frac{\de I_0}{\de\dot c} =\bar c,
\3\Pi_{\Phi}\equiv\frac{\de I_0}{\de\dot \Phi} =\Phib
\eea
so there is no need to introduce independent momenta for these
fields.  There are thus two vanishing momenta
\bea
\Pi_5\equiv  \frac{\de I_0}{\de \dot A_\mu} =0,
\3\> \Psib_5\equiv\frac{\de I_0}{\de\dot\Psi_\mu} =0
\eea
One can then add the kinetic terms for $A_5$ and $\Psi_5$
and constrain them to
zero by means of lagrange multipliers to obtain
\bea\label{uham}
I_0' = \int d^5x \,\big( \dot A_\mu\Pi^\mu + \dot A_5\Pi^5 +
\dot\Psi_\mu\Psib^\mu + \dot\Psi_5\Psib^5
        + \dot c\bar c - \dot\Phi\Phib - \mcl{H}_C
                       - u^1\G_1 - u^2\G_2 \big)
\eea
where
\beq
\G_1\equiv\Psib_5\2\trm{and}\2 \G_2\equiv\Pi_5=s\G_1
\eeq
are the primary constraints and the $u^j$ are new
lagrange multiplier fields.

Now, however, $\{H_C,\G_j\}\approx 0$ generates two secondary
constraints
\beq
\G_3\equiv D_\mu\Psib^\mu + [\Phib,c] -\bar c \2 \trm{and} \2
\G_4 \equiv D_\mu\Pi^\mu + [\Psib_\mu,\Psi^\mu] +[\bar c ,c]
+ [\Phib,\Phi] = s\G_3
\eeq
and these are precisely $\vph_j$ from the minimal treatment.
So the constraints $\vph_1=\G_3$ and $\vph_2=\G_4$ are
now secondary.  We find the following Lie algebra
\beq\label{alg2}\bsp
\{ \G_3^a(x),\G_4^b(y)\} &= \de(x-y)f^{ab}_{\2 \,c} \G_3^c(y) \\
\{ \G_4^a(x),\G_4^b(y)\} &= \de(x-y)f^{ab}_{\2 \,c} \G_4^c(y) \\
               \trm{rest} &= 0
\end{split}\eeq
which is similar to (\ref{alg1}). One finds
$\{H_C,\G_3\}=\{H_C,\G_4\} =0$,
thus there are no tertiary constraints.  As before,
all the constraints are irreducible.

Before we take a look at how the fields actually transform under
gauge transformations, we notice that the constraints $G_m$ may be
replaced by a new set of constraints, given by a linear
combination of the old, $\tilde \G_m = L_m^{\2 n}\G_n$, provided
that the matrix $L$ is invertible, so that the new $\tilde \G_m$
are still irreducible.  Therefore it is permissible to define
\beq
\bsp
\spc{\tilde \G}_1 =\G_1, &\spc {\tilde \G}_3
                \equiv \G_3-[\Psib_5,A_5],\\
\spc{\tilde \G}_2 = \G_2, &\spc {\tilde \G}_4 \equiv \G_4
           +[\Psib_5,\Psi_5] +[A_5,\Pi_5] = s{\tilde \vph}_3
\end{split}
\eeq
The algebra of the ${\tilde \vph}_m$ is slightly
different from (\ref{alg2})
\beq\label{alg3}\bsp
\{ \t \G_1^a(x),\t \G_4^b(y)\} = \de(x-y)f^{ab}_{\2 \,c} \t \G_1^c(y)
   \3&\3 \{ \t \G_2^a(x),\t \G_4^b(y)\} =
                       \de(x-y)f^{ab}_{\2 \,c} \t \G_2^c(y)\\
\{ \t \G_2^a(x),\t \G_3^b(y)\} = \de(x-y)f^{ab}_{\2 \,c} \t \G_1^c(y)
   \3&\3 \{ \t \G_3^a(x),\t \G_4^b(y)\} =
                       \de(x-y)f^{ab}_{\2 \,c} \t \G_3^c(y)\\
\{ \t \G_4^a(x),\t \G_4^b(y)\} = \de(x-y)f^{ab}_{\2 \,c} \t \G_4^c(y)
      \3 &\spc\qquad\;        \trm{rest} = 0
\end{split}\eeq
Note again that the action of $s$ on the left--hand column
gives the right--hand column.
The ${\tilde \G}_3$ and ${\tilde \G}_4$ generate new gauge
transformations
on the $A_5$ quartet, while the rest
of the fields transform the same way.  We list the non--trivial
gauge transformations
\rnc{\dep}[1]{\de_{\veps_{{}_{#1}}}}
\rnc{\tdep}[1]{{\tilde\de}_{\veps_{{}_{#1}}}}
\bal\label{epsalg2}
\dep{1}\Psi_5^a &=\veps_1^a
        & \dep{4}A_\mu^a &=-(D_\mu\veps_4)^a
           & \tdep{3}\Psi_5^a &=[\veps_3,A_5]^a\nn\\
\dep{2}A_5^a &=\veps_2^a
         & \dep{4}\Psi_\mu^a&=[\veps_4,\Psi_\mu]^a
             & \tdep{3}\Pi_5^a &=[\veps_3,\Psib_5]^a\nn\\
\dep{3}\Pi_\mu^a &=[\veps_3,\Psib_\mu]^a
         & \dep{4}\Psib_\mu^a &=[\veps_4,\Psib_\mu]^a
              & \tdep{4}A_5^a &=[\veps_4,A_5]^a \nn\\
\dep{3}\Psi_\mu^a &=-(D_\mu\veps_3)^a
         & \dep{4}\Pi_\mu^a &=[\veps_4,\Pi_\mu]^a
              & \tdep{4}\Psi_5^a &=[\veps_4,\Psi_5]^a\\
\dep{3}\Phi^a &= [\veps_3,c]^a
        & \dep{4}c^a &=[\veps_4,c]^a
              & \tdep{4}\Psib_5^a &=[\veps_4,\Psib_5]^a\nn\\
\dep{3}c^a &=-\veps_3^a
         & \dep{4}\bar c^a &=[\veps_4,\bar c]^a
               & \tdep{4}\Pi_5^a &=[\veps_4,\Pi_5]^a\nn\\
\dep{3}\bar c^a &= [\veps_3,\Phib]^a
         & \dep{4}\Phi^a &=[\veps_4,\Phi]^a
               & {}&{}\nn\\
{}&{}
      & \dep{4}\Phib^a &=[\veps_4,\Phib]^a
               &  {}&{}\nn
\end{align}
If one sets $\veps_1 = -\dot \veps_3$ and $\veps_2 = -\dot\veps_4$
one finds the full gauge transformations to be
\bal
\delta A_\nu&=-D_\nu\veps_4 & \delta\Psi_\nu = [&\veps_4,\Psi_\nu]
-D_\nu\veps_3 &  \delta\Psib_\nu =& [\veps_4,\Psib_\nu]
   & \delta \Pi_\nu =& [\veps_4,\Pi_\nu] + [\veps_3,\Psib_\nu]\nn\\
\delta A_5&=-D_5\veps_4 & \delta\Psi_5 = [&\veps_4,\Psi_5]
-D_5\veps_3 &  \delta\Psib_5 =& [\veps_4,\Psib_5]
   & \delta \Pi_5 =& [\veps_4,\Pi_5] + [\veps_3,\Psib_5]\\
 \delta c = &[\veps_4,c] -\veps_3 & \delta\Phi = [&\veps_4,\Phi]
+ [\veps_3,c] & \delta\Phib =& [\veps_4,\Phib] &
       \delta\bar c =&[\veps_4,\bar c,]  + [\veps_3,\Phib]\nn
\end{align}
which again agrees with the corresponding part of the $w$ algebra
(\ref{BrsAlg}) if the infinitesimal
$\veps$ gauge parameters are replaced by $\lam$ and $\mu$.
\rnc{\dep}[1]{\de_{\eps_{{}_{#1}}}}
\rnc{\tdep}[1]{{\tilde\de}_{\eps_{{}_{#1}}}}

\section{Appendix B: Propagator correction}

In this section we work out in detail the evaluation of the
(transverse) one--loop propagator correction.  The amputated
diagram is given by the integral
\bea
\hat\Sig^{\trm{tr}}(E,p)
 &=& \frac{2}{a^2}\int\frac{d^dk\,d\om}{(2\pi)^{d+1}}\,
     \frac{p^\mu}{\om^2+(k^2)^2}\Big(\de_{\mu\nu}
                              -\frac{k_\mu k_\nu}{k^2}\Big)
     \frac{(p\!+\!k)^\nu}{i(E\!+\!\om)+(p\!+\!k)^2/a}\\
&=&  \frac{2}{a^2}\int\frac{d^d k\,d\om}{(2\pi)^{d+1}}\,
     \big( \de_{\mu\nu}k^2\!-k_\mu k_\nu \big)
                             p^\mu (p\!+\!k)^\nu\nn\\[-10pt]
& &\\[-10pt]
 & &  \bspc \frac{1}{k^2}\,\frac{1}{i\om + k^2}\,\frac{1}{-i\om +k^2}
      \frac{1}{i(E\!+\!\om)+(p\!+\!k)^2/a}\nn
\eea
Using the parameter integral (\ref{ParInt}) this is
expressed as
\bea
& & \frac{2}{a^2}\int\frac{d^d k\,d\om}{(2\pi)^{d+1}}\,
    [ p^2k^2\!-(p\!\cdot\!k)^2]
    \int_0^\infty \!\!d\al\, e^{-[i(E+\om)+(p+k)^2/a]\al}\nn\\[-10pt]
& &\\[-5pt]
 & & \bspc\int_0^\infty \!\!d\be \,e^{-(k^2+i\om)\be}
    \int_0^\infty \!\!d\gam \,e^{-(k^2-i\om)\gam}
    \int_0^\infty \!\!d\lam \,e^{-k^2\lam}\nn
\eea
which after carrying out the $d\om$ integration yields
\bea
& &  \frac{2}{a^2}\int\frac{d^d k}{(2\pi)^d}\,
     [ p^2k^2\!-(p\!\cdot\!k)^2]
    \int_0^\infty \!\!d\al\,\int_0^\infty \!\!d\be\
                  \int_0^\infty \!\!d\lam\,\nn\\[-10pt]
& &\\[-5pt]
& &    \spc \int_0^\infty \!\!d\gam \,\de(\al\!+\!\be\!-\!\gam)
       \exp{-\big[\big((p\!+\!k)^2/a-iE\big)\al+k^2
                    (\be\!+\!\gam\!+\lam)\big]}\nn\\
&=&  \frac{2}{a^2}\int\frac{d^d k}{(2\pi)^d}\,
     [ p^2k^2\!-(p\!\cdot\!k)^2]
    \int_0^\infty \!\!d\al\,\int_0^\infty \!\!d\be\
              \int_0^\infty \!\!d\lam\,\nn\\[-10pt]
& &\\[-5pt]
 & &  \bspc\exp{-\big[(k^2+2k\!\cdot\!p + p^2)\al/a
              -iE\al + k^2(\lam\!+\!\al\!+\!2\be)\big]}\nn
\eea

\flush Changing variables $\al\ra\al/a$, $\be\ra\be/2$ gives
 \bea
\<\int\frac{d^d k}{(2\pi)^d}\,[ p^2k^2\!-(p\!\cdot\!k)^2]
    \int_0^\infty \!\!d\al\,\int_0^\infty
                               \!\!d\be \int_0^\infty \!\!d\lam\,
    \exp{-\big[\xi k^2 \!+2(k\!\cdot\!p)\al +p^2\al -iEa\al\big]}
\eea
where we defined $\xi=\al+a(\al\!+\!\be)+\lam$.
Shifting $k\ra k+\al p/\xi$ and noting that
$p^2k^2\!-(p\!\cdot\!k)^2$ is translation--invariant,
this becomes
\bea
\int_0^\infty \!\!d\al\,\int_0^\infty \!\!d\be \int_0^\infty \!\!d\lam\,
e^{-\al\big(\frac{\xi-\al}{\xi}p^2+iEa \big)}
\int\frac{d^d k}{(2\pi)^d}\,[ p^2k^2\!-(p\!\cdot\!k)^2]e^{-\xi k^2}
\eea

Performing the $d^dk$ integration (we take the symmetric limit
$k^\mu k^\nu\ra \delta^{\mu\nu}k^2/d)$ gives
\bea
\frac{(d\!-\!1)p^2}{d(16\pi)^{d/2}}
\int_0^\infty \!\!d\al\,\int_0^\infty \!\!d\be \int_0^\infty \!\!d\lam\,
\frac{1}{\xi^{d/2+1}}\,e^{-\al\big(\frac{\xi-\al}{\xi}p^2+iEa \big)}
\eea
We next insert $1=\displaystyle\int_0^\infty d\eta\,\de(\eta\!-\!\xi)$
into the integral and change variables
\bea
\al\ra\eta\al,\2\be\ra\eta\be,\2\lam\ra\eta\lam\3\Ra\3\xi\ra\eta\xi
\eea
noting that the delta function transforms as
$\de(\eta\!-\!\xi)\ra\de(\eta(1\!-\!\xi))=\de(\xi\!-\!1)/\eta$
and the integral becomes
\bea
C(d)p^2
\int_0^\infty \!\!d\al\,\int_0^\infty \!\!d\be
\int_0^\infty \!\!d\lam\,\de(\xi\!-\!1)
\int_0^\infty \frac{d\eta}{\eta}\,\eta^{\frac{4-d}{2}}
e^{-\al[(1-\al)p^2+iEa]\eta}
\eea
We have defined
\bea
C(d)\equiv \frac{d\!-\!1}{d(16\pi)^{d/2}}
\eea
to unburden the notation.  Observing that the delta function
$\de(\xi\!-\!1)=\de(\lam+(1\!+\!a)\al+a\be-1)$ effectively constrains
the $d\al\, d\be\, d\lam$ integration to a 2--simplex, we integrate
out $d\lam$ to get
\bea
C(d)p^2\<\<
\underset{\al,\be\vgeq 0}{\underset{0\vleq(1+a)\al+a\be\vleq 1}
{\int\!\! \int}}
 \!\<\<d\al\,d\be\,\, \int_0^\infty \frac{d\eta}{\eta}\,
\eta^{\frac{4-d}{2}} e^{-\al[(1-\al)p^2+iEa]\eta}
\eea
Integrating over $d\be$ and $d\eta$ results in
\bea
\hat\Sig^{\trm{tr}}(E,p)=
\frac{C(d)}{a}p^2
\int_0^\frac{1}{1+a}\!\!d\al\,[1\!-\!(1\!+\!a)\al]
\frac{\Gam(\eps)}{\al^\eps [(1\!-\!\al)p^2+iEa]^\eps}
\eea
where $\eps\equiv(4-d)/2$ is the dimensional regularization parameter.
Attaching the legs onto $\hat\Sig$
and taking the inverse Fourier transform in $E$
gives the one loop correction
\bea\label{D1}
\hat{\mcl{D}}^{\trm{tr}}_1(t,p)
&=& (\hat {\mcl{D}}_0 \hat\Sig^{\trm{tr}}
                            \hat{\mcl{D}}_0)^{\vee}(t,p)^{}\nn\\
&=& \int\frac{dE}{2\pi}\,e^{iEt}\,\hat{\mcl{D}}^{\trm{tr}}_1(E,p)\\
&=&\Gam(\eps)\frac{C(d)}{a}p^2
\int_0^\frac{1}{1+a}\frac{d\al}{\al^\eps}\,[1\!-\!(1\!+\!a)\al]
\int\frac{dE}{2\pi}\,\frac{e^{iEt}}
              {[(1\!-\!\al)p^2+iEa]^\eps(iE+p^2/a)^2}\nn
\eea
Scaling $E\ra E/a$ leads to
\bea
\<\<\hat{\mcl{D}}^{\trm{tr}}_1(t,p)
=\Gam(\eps)C(d)p^2
\int_0^\frac{1}{1+a}\frac{d\al}{\al^\eps}\,[1\!-\!(1\!+\!a)\al]
\int\frac{dE}{2\pi}\,\frac{e^{iEt/a}}
              {[(1\!-\!\al)p^2+iE]^\eps(p^2+iE)^2}
\eea

Let us now deal with the $dE$ integral
\bea
\int\frac{dE}{2\pi}\,\frac{e^{iEt/a}}
              {[(1\!-\!\al)p^2+iE]^\eps(p^2+iE)^2}
\eea
Rewriting the denominator using the parameter integrals, this equals
\bea
\int \frac{dE}{2\pi} \,e^{iEt/a}\frac{1}{\Gam(\eps)}
\int_0^\infty\frac{dx}{x}\, x^\eps e^{-[(1\!-\!\al)p^2+iE]x}
\int_0^\infty dy\,y\,e^{-(p^2+iE)y}
\eea
Integrating out $dE$ gives a $\de$--function so we have
\bea
\frac{1}{\Gam(\eps)}\int_0^\infty\frac{dx}{x} \,x^\eps
\int_0^\infty dy\,y\,\de(x\!+\!y\!-\!t/a) e^{-[(1-\al)x+y]p^2}
\eea
Letting $x\ra xt/a$, $y\ra yt/a$ (and so $\de(x\!+\!y\!-\!t/a)\ra
a\de(x\!+\!y\!-\!1)/t$) gives
\bea
& &\frac{1}{\Gam(\eps)}\Big(\frac{t}{a}\Big)^{1+\eps}
\int_0^\infty\frac{dx}{x} \,x^\eps
\int_0^\infty dy\,y\,\de(x\!+\!y\!-\!1) e^{-[(1\!-\!\al)x+y]p^2t/a}\\
&=&\frac{1}{\Gam(\eps)}\Big(\frac{t}{a}\Big)^{1+\eps}
\int_0^1\frac{dx}{x} \,x^\eps (1\!-\!x)e^{(\al x-1)p^2t/a}
\eea
So we obtain the following expression for the one--loop
correction (\ref{D1}) (note that $\Gam(\eps)$ cancels)
\bea
\hat{\mcl{D}}^{\trm{tr}}_1(t,p) =
C(d)p^2
e^{-p^2t/a}\Big(\frac{t}{a}\Big)^{1+\eps}
\int_0^\frac{1}{1+a}\frac{d\al}{\al^\eps}\,\big(1\!-\!(1\!+\!a)\al\big)
\int_0^1\frac{dx}{x} \,x^\eps(1\!-\!x) e^{\al x p^2t/a}
\eea
which becomes
\bea
\frac{C(d)}{1\!+\!a} \Big(\frac{1\!+\!a}{a}\,t\Big)^\eps
e^{-p^2t/a}\big(p^2t/a\big)
\int_0^1\frac{d\al}{\al^\eps}\,(1\!-\!\al)
\int_0^1\frac{dx}{x} \,x^\eps(1\!-\!x) e^{-\al\frac{p^2t}{a(1+a)}x}
\eea
after rescaling $\al\ra\al/(1\!+\!a)$.  Clearly this integral is
finite for $\eps>0$.  We Taylor expand the integrand in $t$ and
integrate $dx$ to find
\bea
\frac{C(d)}{1\!+\!a} \Big(\frac{1\!+\!a}{a}\,t\Big)^\eps
e^{-p^2t/a}\big(p^2t/a\big)
\int_0^1\frac{d\al}{\al^\eps}\,(1\!-\!\al)
\Big[\frac{1}{\eps}-1 + \frac{\al p^2t}{2a(1\!+\!a)} +O(\eps)\Big]
\eea
The remainder $O(\eps)$ consists of terms constant and higher
order in $t$.  Evaluating the remaining $d\al$ integral
we finally have
\bea
\hat{\mcl{D}}^{\trm{tr}}_1(t,p)
&=& \frac{C(4)}{2(1\!+\!a)} \Big(\frac{1\!+\!a}{a}\,t\Big)^\eps
e^{-p^2t/a}\big(p^2t/a\big)\Big[
\frac{1}{\eps}+1 + \frac{p^2t}{6a(1\!+\!a)} +\cdots  \Big]\\
&=&\frac{3/2}{(32\pi)^2}
e^{-p^2t/a}\frac{p^2t}{a(1\!+\!a)}\Big[
\frac{1}{\eps}+1-\log\frac{t}{a(1\!+\!a)} +O(\eps\log t)\Big]
\eea
which is proportional to $t$. Apparently the $1/\eps$ pole generates a
time renormalization counterterm, but we will not address the
details of renormalizing the theory here.  What is important is
that the correction vanishes as $t\sea 0$.


\begin{thebibliography}{99}

\bibitem{Pesk}
Michael E. Peskin, Daniel E. Schroeder
{\em An Introduction to Quantum Field Theory} Perseus Books 1995
\newblock


\bibitem{BaZwQCD}
Laurent Baulieu and Daniel Zwanziger,
{\em{QCD}$_4$ From a Five-Dimensional Point of View},
\newblock Nucl. Phys. B 581 (2000) 604 hep-th/9909006


\bibitem{BaZwScal}
Laurent Baulieu and Daniel Zwanziger,
{\em From stochastic quantization to bulk quantization: Schwinger-Dyson
equations and S-matrix},
\newblock  JHEP 08:016 (2001) hep-th/0012103


\bibitem{BaGrZw}
Laurent Baulieu, Pietro~Antonio Grassi, and Daniel Zwanziger,
\newblock {\em Gauge and Topological Symmetries in the Bulk
Quantization of Gauge Theories},
\newblock  Nucl. Phys. B 597 (2001) 583,   hep-th/0006036
\newblock

\bibitem{BaZwGauge}
Laurent Baulieu and Daniel Zwanziger,
{\it Bulk Quantization of Gauge Theories: Confined and Higgs Phases},
\newblock JHEP 08:015 (2001) hep-th/0107074
\newblock

\bibitem{Stoch}
P.H. Damgaard and H. H\"uffel Eds., {\em Stochastoc Quantization},
World Scientific 1988
\newblock


\bibitem{Stoch2}
M. Namiki and K. Okano Eds., {\em Stochastic Quantization}
Prog. Theor. Phys. Suppt 111 (1993)
\newblock


\bibitem{Bert}
Reinhold Bertlmann,
{\em Anomalies in Quantum Field Theory}, Oxford University Press 1996
\newblock


\bibitem{Bau}
Laurent Baulieu,
{\em Perturbative Gauge Theories}, Phys.Rep. 129 (1985) 1
\newblock

\bibitem{Sund}
Kurt Sundermeyer,
{\it Constrained Dynamics}, Springer Verlag 1982
\newblock


\bibitem{HenTeit}
Marc Henneaux and Claudio Teitelboim,
{\it Quantization of Gauge systems},
Princeton University Press 1992
\newblock

\bibitem{Bat}
Batalin et. al., Phys.Lett. B69 (1977) 309
\newblock

\bibitem{Hen}
Marc Henneaux, Phys.Rep. 126 (1985) 1
\newblock

\bibitem{PDE1}
Michael E. Taylor,
{\it Partial Differential Equations I: Basic Theory},
Springer Verlag 1996
\newblock



\end{thebibliography}
\end{document}